\documentclass{aa}
\input{psfig.sty}
\begin{document}
\title{Timing noise, glitches and the braking index of PSR~B0540$-$69}
\author{G. Cusumano\inst{1}, E. Massaro\inst{2,3}, T. Mineo\inst{1}
}

\institute{Istituto di Astrofisica Spaziale e Fisica Cosmica - Sezione di Palermo, CNR, 
Via Ugo La Malfa 153, I-90146, Palermo,
Italy \and
Dipartimento di Fisica, Universit\'a La Sapienza, Piazzale A. Moro 2,
I-00185, Roma, Italy \and
Istituto di Astrofisica Spaziale e Fisica Cosmica - Sezione di Roma, CNR,
Via del Fosso del Cavaliere, I-00100 Roma, Italy 
}

\offprints{G. Cusumano: giancarlo.cusumano@pa.iasf.cnr.it}
\date{Received:.; accepted:.}

\titlerunning{Timing, glitches and braking index of PSR~B0540$-$69}
\authorrunning{G. Cusumano et al.}

\abstract{ We report a pulse-time history of PSR~B0540$-$69 based on the
analysis of an extended Data set including  ASCA, BeppoSAX and RXTE 
observations spanning a time interval of about 8 years. 
This interval includes also the epoch of the glitch episode reported 
by Zhang et al. (2001). Our analysis shows the presence of a relevant 
timing noise and does not give a clear evidence of the  glitch occurrence. 
We performed an accurate evaluation of the main timing parameters, $\nu$, 
$\dot{\nu}$ and $\ddot{\nu}$  and derived a mean braking index  of 
$n=2.125\pm0.001$  quite different from the lower value found by Zhang et al. 
(2001), but in rather good agreement with other several values reported 
in the literature.
}

\maketitle

\section{Introduction}
PSR B0540$-$69 was discovered in the soft X-rays by Seward et al. (1984) 
in a field of the Large Magellanic Cloud observed with the Einstein
Observatory.
Pulsations at optical frequencies were soon detected by Middleditch \&
Pennypacker (1985) with a mean pulsed magnitude of 22.5.  
In the radio band PSR B0540$-$69  is a quite faint source and pulsed signals  
were first observed only in 1989-90 (Manchester et al. 1993).

PSR B0540$-$69 is one of the youngest rotation powered pulsars. It has a period 
of about 50 ms and a large period derivative of 
$4.79\times10^{-13}$ s
 s$^{-1}$,
comparable to that of the Crab pulsar, and a spin down age of about 1500 years.
The pulse shape, at X and optical wavelengths, is broad  and almost sinusoidal. 
Several estimates of the braking index $n$ have been reported in the
literature (see Boyd et al. 1995 for a compilation of older results)
ranging from $2.01\pm0.02$ (Manchester \& Peterson 1989; Nagase et al. 1990)
to $2.74\pm0.10$ (\"{O}gelman \& Hasinger 1990). 
Deeter et al. (1998) analyzed an extended set of GINGA 
observations and found a braking index of 2.08$\pm$0.02. This result was
substantially confirmed by Mineo et al. (1999), who combining a BeppoSAX frequency 
measurement with earlier ASCA results derived a value of 2.10$\pm$0.1. 
A recent analysis of all ASCA pointed observations gave a braking 
index of 2.10 (Hirayama et al. 2002).
A glitch in the timing history of PSR~B0540$-$69, the only one reported up to now
for this young Crab-like pulsar, has been recently pointed out by
Zhang et al. (2001). 
These authors based their analysis on a collection of RXTE observations, 
spanning 1.2 years. 
The approximate epoch of the glitch, according to their estimates,
occurred on MJD 51325 $\pm$ 45 and the change in frequency and its first derivative were
$\Delta\nu / \nu = (1.90 \pm 0.04) \times 10^{-9}$ and $\Delta\dot{\nu} / \dot{\nu} = 
(8.5 \pm 0.07)
\times 10^{-5}$, respectively. Zhang et al. (2001) measured also the 
braking index after the glitch which resulted equal to 1.81 $\pm$ 0.07, 
significantly lower than the values reported from other analyses.

In this paper we present an exhaustive timing analysis of PSR~B0540$-$69
performed on RXTE data set spanning aboUt 5 years. 
In particular, we extend the set of RXTE observations used by Zhang et al. 
(2001), adding more observations for a total time interval of about 3 years 
before and 2 years after the epoch of the glitch episode claimed by
these authors. Moreover, to further increase the time 
interval data set, public ASCA and BeppoSAX observations are also 
included in the analysis extending the length up to about 8 years.

\section{Observation and Data Reduction}
The RXTE observations considered in the present paper were performed between 
March 8, 1996 and March 14, 2001.
We used only data obtained with the PCA (Jahoda et al. 1996) 
accumulated in ``Good Xenon'' telemetry mode, time-tagged with a 
1$\mu$s accuracy with respect to the spacecraft clock, which is maintained 
to the UTC better than 100 $\mu$s. 
The pulsar position inside the instrument Field of View (FoV) is different in
the various pointings and ranges 
between 0 and 25 arcminutes far from the centre.
Following Zhang et al. (2001) we considered only events with pulse-height channels 
between 5 and 50, corresponding to the 2-18 keV energy interval.
Furthermore, we verified that the selection of all PCA detector layers, instead of 
those from the top layer only, increased significantly the S/N ratio of the pulsation 
and adopted this choice for all the observations.

PSR~B0540-69 was observed by ASCA (Tanaka et al. 1994) 14 times between 
1993-06-13 and 1999-11-03.
Only data from GIS (Ohashi et al. 1996, Makishima et al, 1996) were  used in our 
analysis.
The pulsar position inside the FoV ranges between 0 and 22 arcminutes 
off-axis. The Narrow Field Instruments (NFIs) onboard BeppoSAX (Boella et al. 1997a) 
observed PSR~B0540-69 five times between 1996-10-26 and 2000-02-19. 
We considered only data from MECS (Boella et al. 1996) because they are in 
the same energy range used for the analysis of the other two satellites.
The position of the source inside the instrument FoV lies between 0 and 17 arcminutes. 
Data from both imaging instruments  (GIS and MECS) were extracted from regions  
centered at the source position and the shapes and sizes of these regions were 
optimized for each observation, depending upon the different off-axis locations, 
to achieve the highest S/N ratio.
The log of all these observations is given in Table A1, available in electronic form
at CDS (http://cdsweb.u-strasbg.fr/).

\section{Timing analysis and pulse profile}

\subsection{Frequencies and derivatives evaluation}
The UTC arrival times of all selected events were first converted to
the Solar System 
Barycentre using the (J2000) pulsar position $\alpha = 5^{\rm h} 40^{\rm m} 
10\fs95^{\rm }$ and $\delta = -69
\degr 19 \arcmin 55\farcs 1$ (Caraveo et al. 1992) 
and the JPL2000 ephemeris (DE200; Standish 1982).
For each observation we searched the pulsed frequency
$\nu$ using the folding 
technique in a range centered at the values
computed by means of the 
pulsar ephemeris reported by Deeter et al. (1999).
The central time of each observation was chosen as reference epoch and the 
corresponding frequency was estimated by fitting the $\chi^2$ peak with a gaussian
profile. Both these data are listed in Table A1 (columns 2 and 7, respectively).
Frequency errors at 1 $\sigma$ level were computed with two different methods.
The first evaluation was performed computing the frequency interval corresponding 
to a unit decrement with respect to the maximum in the $\chi^2$ curve 
($err=\nu(\chi^2_{max})-\nu(\chi^2_{max}-1)$).
In the second method we produced a template profile folding the longest RXTE observation 
(rows 33$+$34 in Table A1) with its best frequency (see Fig.1). 
The zero phase was taken at the centre of the peak.
Then, we compared the folded profiles of other pointings
with an accurate analytical 
model of the template 
changing the frequency within the searching range and computed the relative $\chi^2$ 
values. 
For a correct comparison we subtracted the off-pulse levels and normalized the total pulsed counts.
The frequency uncertainties were then set equal to the frequency interval corresponding 
to an increase of the $\chi^2$ by a unit with respect to the minimum in the curve 
($err=\nu(\chi^2_{min}+1)-\nu(\chi^2_{min})$).
No significant difference was found between the error estimates obtained with the two 
methods, we then considered as proper statistical uncertainties of
the pulsar 
frequencies those computed with the first one (column 8). 
All errors, thereafter, refer to one standard deviation. 
The times of arrival (TOA), referred to the peak centre, were also determined for 
all RXTE observations by cross-correlating the folded profiles with the analytical 
model of the template. The resulting phase offset was added to the central 
time of the observation. Errors on TOAs were assumed equal to the statistical uncertainties
of the peak centre derived from a best fit of the pulse profile. TOA and  errors are reported 
in Table A1 (columns 5 and 6.) 

\begin{figure}
\centerline{
\vbox{
\psfig{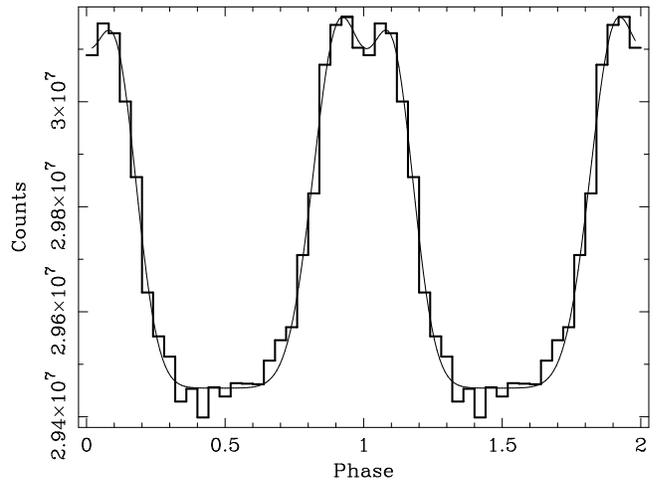}
}}
\caption{The (2--18 keV) pulse profile of PSR~B0540$-$69 in 25 phase bins.
The analytical model used as
template for the timing analysis is also shown.
}
\end{figure}

A first estimate of the frequency derivatives was obtained by the best fit of the 
frequency history listed in Table A1 with the second-degree polynomial: 

\begin{equation}
\nu(t) = \nu_0 + \dot{\nu}_0 (t-t_0) + \frac{1}{2}\ddot{\nu}_0(t-t_0)^2\;.
\end{equation}

This procedure was first applied separately to the two ASCA+BeppoSAX and RXTE frequency 
sets and the resulting parameters' values were found in a good agreement within their 
uncertainties. 
We then evaluated the ephemeris for the whole set of data (RXTE+ASCA+BeppoSAX); a summary
of these results is given in Table 1. 
For easy comparison, the same reference epoch ($t_0$) was taken, in all the fittings, 
equal to MJD 50372.5481748585. 
Fig. 2 shows the residuals obtained fitting the entire frequency set with Eq.(1): 
the number of frequency values in excess of 2 standard deviations is about
6\% of the total, confirming the right evaluation of the frequency errors.

\begin{figure}
\centerline{
\vbox{
\psfig{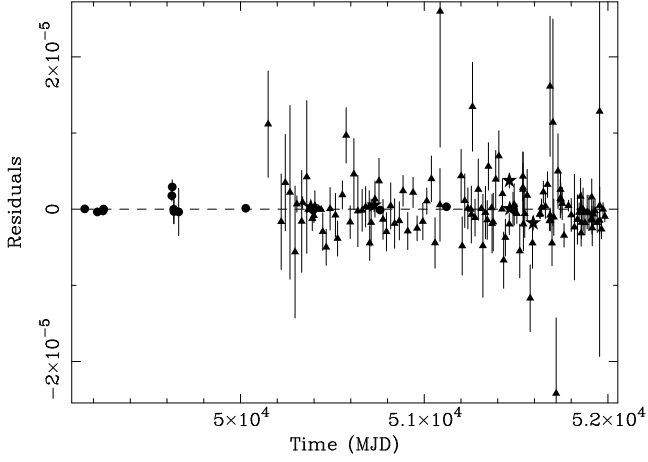}
}}
\caption{Residuals of the best fit of the frequencies obtained by 
the folding technique with the polinomial formula of Eq.(1). Triangles represents RXTE 
data, circle ASCA data and stars BeppoSAX data.
}
\end{figure}

\begin{figure}
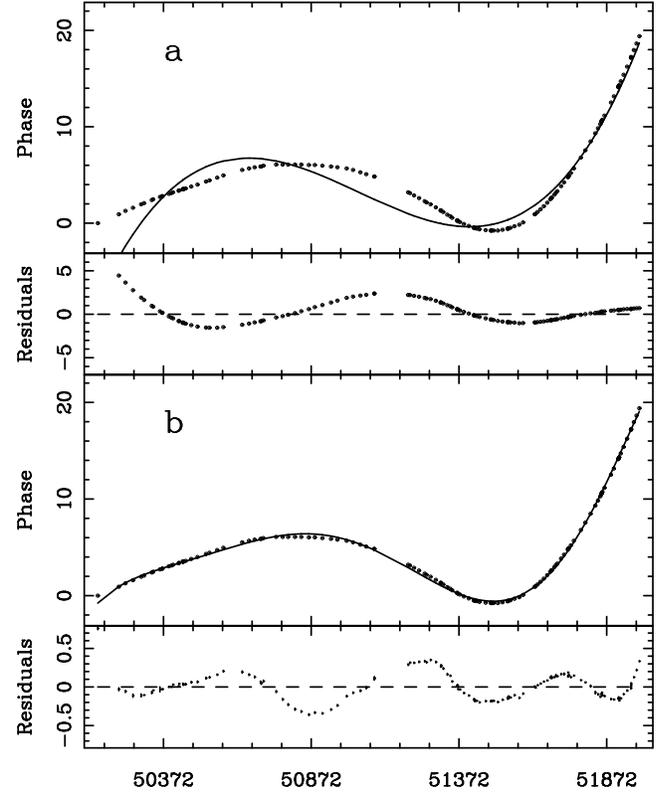

\centerline{
\vbox{
\psfig{figure=figura3a.ps,width=8.5cm,angle=-90}
\vspace{-0.1 cm}
\psfig{figure=figura3b.ps,width=8.5cm,angle=-90}
}}
\caption{Phase of the pulsed signal, after a careful reconstruction
 of the 
phase recycles for the entire set of the RXTE observations, vs observing time.
Phase is in cycle 
units.
Panel $a$ shows the fit with a
third order polynomial and panel $b$ shows the fit with a sixth order polynomial. 
Data and fitting models
are shown in the top, residuals in the bottom. 
}
\end{figure}

\begin{table*}
\caption{Ephemerides of PSR~B0540$-$69 obtained from a second-degree polynomial fit of the frequency 
data sets reported in Table A1. 1 $\sigma$ error is given in parentheses 
for the corresponding last significant digits.}
\begin{tabular}{lllll}
\hline
\hline
\multicolumn{1}{c}{Data Set} &\multicolumn{1}{c}{t$_0$} & \multicolumn{1}{c}{$\nu$} & \multicolumn{1}{c}{$\dot{\nu}$ } & \multicolumn{1}{c}{$\ddot{\nu}$}   \\
         &\multicolumn{1}{c}{(MJD)}  & \multicolumn{1}{c}{(Hz)}   & ($\times 10^{-10}$ Hz $s^{-1}$) & ($\times 10^{-21}$ Hz $s^{-2}$)  \\
\hline
ASCA+BeppoSAX      &50372.5481748585 &19.81583115(7) &\hspace{.4cm}-1.880724(7)& \hspace{.4cm} 3.74(2) \\
RXTE               &50372.5481748585 &19.8158314(1)  &\hspace{.4cm}-1.8808(1)& \hspace{.4cm} 3.73(20)  \\
ASCA+BeppoSAX+RXTE &50372.5481748585 &19.81583119(4) &\hspace{.4cm}-1.880727(8)& \hspace{.4cm} 3.717(7) \\
\hline
\end{tabular}
\end{table*}

\subsection{Pulse phase analysis}
The precision of the timing parameters can be enhanced by maintaining a 
pulse coherence over the entire time interval covered by the observations. 
This analysis was performed only on the RXTE data.  The BeppoSAX timing,
in fact, does not maintain the indispensable accuracy to the UTC,
fundamental to apply a coherent phase analysis. The ASCA observations are 
rather sparse and the systematics in the absolute time assignment 
(Hirayama et al. 1996) makes more complicated the proper 
phase recycle correction.

The arrival times of every event for RXTE observations were folded by using the values 
of $\nu_o$, $\dot{\nu}_o$, $\ddot{\nu}_o$ reported in Table 1 (line 3).
Phase shifts of the pulse profiles, expected because of the ephemeris 
accuracy, were computed for all observations by a cross-correlation 
with the analytical model of the template (Fig. 1). 
Phase errors were taken equal to the TOA uncertainties multiplied by the corresponding 
frequencies. Throughout this paper phases are measured in cycle units.
The resulting values of the phase shifts, corrected for the presence of recycles, are 
shown in Fig. 3 as a function of the observation epoch. 

The large phase variations, of the order of several cycles in a few hundred days,
indicate that input ephemerides of Table 1 are not the best ones and that it is possible 
to improve them. The corrections, in absence of frequency irregularities, can be
obtained by fitting the phase lags by the simple third-degree spin-down 
model:

\begin{equation}
\Delta \phi(t)=\Delta\nu_o(t-t_0) + \frac{1}{2}\Delta\dot
{\nu}_o (t-t_0)^2 + \frac{1}{6}\Delta\ddot{\nu}_o (t-t_0)^3\,\,,
\end{equation}

\noindent 
where $\Delta \phi(t)$ is the measured phase difference at time $t$, $t_0$
is the reference epoch and $\Delta\nu_o$, $\Delta\dot{\nu}_o$, $\Delta\ddot{\nu}_o$ 
are the ephemeris corrections for the frequency, first and second derivatives, 
respectively. 
We found that this equation fails to describe the entire set of 
phase data showing very large systematic deviations (panel $a$ of Fig. 3). 
Such deviations indicate the presence of a timing noise 
characteristic of several other pulsars (Arzoumanian et al. 1994, Lyne 1999).
We tried also to correct the phases by fitting higher order polynomials, up to the sixth 
degree. 
The best fit for this model is shown in the panel $b$ of Fig. 3: the residuals are of 
course largely reduced with respect to the other fit and the general behaviour is 
much better described, but smaller amplitude systematic deviations remain apparent.

To quantify the timing noise we can calculate the $\Delta_8$ parameter defined by 
Arzoumanian et al. (1994) as:

\begin{equation}
\Delta_8 = {\rm log} (10^{24} |\ddot{\nu}|/6\nu)\,\,.
\end{equation}

\noindent
For PSR~0540$-$69 instead of the measured $\ddot{\nu}$, whose large 
value is related to the regular spin down rather than the timing noise, 
we used the correction that takes
into account the residuals shown in Fig. 
3a. A reliable estimate of this correction is given by the difference between 
the third degree coefficient of the best fit
parameters of the sixth and third 
degree polynomials. 
This difference
was found equal to $2.27 \times 10^{-22}$ Hz s$^{-2}$ and 
gives $\Delta_8$=0.28,
higher than all the values of the sample of Arzoumanian 
et al. (1994). 
Note that PSR~B0540-69 lies well in the upper
part of the plot $\Delta_8$ 
$vs$ $\dot{P}$ given by these authors. 

For comparing our results with those of Zhang et al. (2001) we divided
our data into two subsets, before and after the time of the glitch,
assumed at MJD 51325 -- the interval of possible epochs given by Zhang et al.
(2001), spans 45 days -- and applied a cubic polynomial fit to each subset. 
We obtained acceptable fits, characterized by nearly zero residuals;
resulting best fit values for the two intervals are reported in Table 2,
the ephemerides are given in Table 3.
Propagating the parameters of Table 3 to the epoch of the glitch we obtained 
a marginal detection for the frequency change 
$\Delta\nu/\nu$ = (1.8$\pm$1.0) $\times$ $10^{-9}$, whereas more significant 
differences were found for the first and second derivative:
$\Delta\dot{\nu}/\dot{\nu}$ =  (1.69$\pm$0.01) $\times$ $10^{-4}$ and
$\Delta\ddot{\nu}/\ddot{\nu}$ =  (2.043$\pm$0.001) $\times$ $10^{-3}$.
The significance of these parameters' differences can be
largely affected by the timing noise systematics. To verify this hypothesis 
we performed a similar analysis on other data subsets selected by  
changing the separation epochs and values of $\Delta\nu/\nu$ of the same 
order and similar significance were found.
This last finding addressed us to interpret the marginal detection of frequency jump 
at the glitch epoch claimed by Zhang et al. (2001) as a not genuine result
but due to the pulse phase analysis in presence of strong timing noise.

\subsection{The braking index}

As discussed in the Introduction, a relevant difference between our previous results 
(Mineo et al. 1999) and those of Zhang et al. (2001) is in the value of the braking 
index: we essentially confirmed the estimates of Deeter et al. (1998), Hirayama et al. (2002), 
while Zhang et al. (2001) gave the value of 1.81, about 14\% smaller, 
but significantly different when considering the associated uncertainties.
Using the new ephemerides given in Table 3, we computed for the two intervals the
values of $n$ and obtained 2.1272$\pm$0.0003 and 2.122$\pm$0.001  for the first and second
interval, respectively.

A change in the timing parameters implies a variation of the braking index:

$$
\frac{\Delta n}{n} =\frac{\Delta\nu}{\nu} - 2 \frac{\Delta\dot{\nu}}{\dot{\nu}} + 
\frac{\Delta\ddot{\nu}}{\ddot{\nu}}  \,\,.
\eqno(4)
$$ 

From the above results it is clear that the parameter responsible for the much lower
value of $n$ measured by Zhang et al. (2001) is the second frequency derivative, 
for which these authors reported the value of (3.23$\pm$0.12)$\times 10^{-21}$ s$^{-2}$, 
just 15\% smaller than our result.

This discrepancy is likely due to the presence of such high timing noise,
because the data set considered by Zhang et al. (2001) spans a rather narrow
time interval with respect to that considered by us. 
To further investigate this point we computed the braking index in several 
independent time intervals, evaluating the best corrections to the pulsar ephemerides in
each of them.
The resulting values are shown in Fig. 4: they ranges from 1.97 to 2.47. 
Again the largest variation is due to the estimates of second derivative of the pulsar
frequency, which differs of $\sim$20\%, between the various intervals.
We conclude, that reliable estimates of the braking index can be obtained only
considering the longest time intervals in which a good fit of phases (or TOA) with Eq.(2) 
can be obtained, likely spanning several years. The use of shorter intervals
can introduce a bias due to the timing noise.

\begin{figure}
\centerline{
\vbox{
\psfig{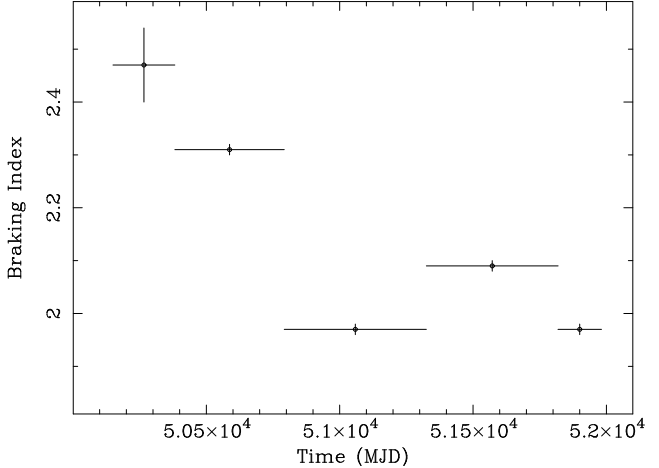}
}}
\caption{The values of the braking index computed in several intervals
plotted as a function of the central time of the interval These large
changes are a consequence of the timing noise affecting the second
derivative of the pulsar frequency. 
}
\end{figure}

\begin{table*}
\caption{Corrections to the timing parameters of PSR~B0540$-$69.
}
\begin{tabular}{llll}
\hline
\hline
\multicolumn{1}{c}{Time Interval}&\multicolumn{1}{c}{ $\Delta\nu$} & \multicolumn{1}{c}{$\Delta\dot{\nu}$} & \multicolumn{1}{c}{$\Delta \ddot{\nu}$ }\\
                             &\multicolumn{1}{c}{ (Hz)}        & \multicolumn{1}{c}{(Hz s$^{-1}$)}     & \multicolumn{1}{c}{ (Hz s$^{-2}$) }\\
\hline
(1) t$<$ 51325 MJD      & -1.359(2)$\times$10$^{-7}$ & 1.68(2)$\times$10$^{-15}$& 8.00(4)$\times$10$^{-23}$\\
(2) t$>$ 51325 MJD      &  2.487(9)$\times$10$^{-6}$   & -2.94(2)$\times$10$^{-14}$& 7.2(2)$\times$10$^{-23}$\\
\hline
\end{tabular}
\end{table*}

\begin{table*}
\caption{Frequency ephemerides obtained from a phase coherent analysis before and after MJD 51325}
\begin{tabular}{llllll}
\hline
\hline
\multicolumn{1}{c}{Time Interval}& \multicolumn{1}{c}{t$_0$}&\multicolumn{1}{c}{$\nu$ } & \multicolumn{1}{c}{$\dot{\nu}$ } & \multicolumn{1}{c}{$\ddot{\nu}$ } & \multicolumn{1}{c}{$n$ }\\
                                 & \multicolumn{1}{c}{(MJD)}&\multicolumn{1}{c}{(Hz)} & \multicolumn{1}{c}{($\times 10^{-10}$ Hz s$^{-1}$)} &\multicolumn{1}{c}{($\times 10^{-21}$ Hz s$^{-2}$)} &  \\
\hline
(1) t$<$ 51325 MJD & 50372.5481748585 & 19.8158310541(2) &\hspace{0.2cm} -1.8807101(2) & \hspace{0.4cm}3.7970(4) & 2.1272(3)\\
(2) t$>$ 51325 MJD & 50372.5481748585 & 19.815833677(9) & \hspace{0.2cm} -1.881021(2) & \hspace{0.4cm}3.789(2) & 2.122(1)\\
\hline
\end{tabular}
\end{table*}

\section{Discussion}

The only way to study the timing noise of PSR~B0540-69 is the use of a dense set
of X-ray observations, like that of RXTE, because this young pulsar is in the Large 
Magellanic Cloud and its flux is too weak to be monitored in the radio band.
Our analysis on a large database of X-ray observations of PSR~B0540-69, covering
more than 5 years, provided a good evidence for a relevant timing noise affecting the 
phase of the pulsed signal. In particular, we showed that the best fit of a third
degree polynomial, including up to the second frequency derivative, gives phase 
residuals up to a few cycles and that residuals as large as 0.4 remain even when a sixth
degree polynomial is used. Assuming that the difference in the second derivative
obtained from these best fits is a measure of its fluctuations, we evaluated the 
$\Delta_8$ parameter (Arzoumanian et al.1994) which was found equal to 0.28, confirming 
the high level of timing noise. Taking into account that even larger variations of this 
derivative are also found when polynomial best fits are performed over shorter time 
intervals, this result could be considered a lower limit. Furthermore, it supports 
the finding that timing noise is stronger in young pulsars with a high $\dot{P}$.
  
A consequence of this high timing noise is that the glitch claimed by 
Zhang et al. (2001)
cannot be confirmed. The frequency difference of this event given by 
these authors is very  small, about 0.04 $\mu$Hz. 
The glitch cannot be detected directly as a sudden frequency jump 
because it is quite less than the uncertainties in the frequency  
measurements which are typically of the order of a few $\mu$Hz, and 
in the best conditions 
of a few tenth of $\mu$Hz.
Zhang et al. (2001) also excluded that this effect can be due 
to timing noise, but their conclusion is affected by the use of a shorter time 
interval that does not allows an accurate analysis of the timing noise.

At variance, our results show that frequency differences 
of the same order of that given
by Zhang et al. (2001)
are usually found when different selections of time intervals
are considered and they do not depend upon a well defined episode. 

From the timing noise analysis we were also able to show how much the first and
second derivative of the pulsar frequency are stable in time. We found that the
former can have fluctuations of amplitude of about 10$^{-4}$, 
while for the latter 
fluctuations can be much higher, and in some cases the estimates can differ of 
$\sim$ 10--20\%, depending upon the length of the time interval taken 
into account.
Such large variations affect the evaluation of the braking index, 
particularly when
it is found by the fitting of Eq.(2) to the pulse phases.
We showed that when the longest possible intervals are considered, $n$ turns out to be
very close to 2.12, in agreement with several other previous estimates, while values
as that given by Zhang et al. (2001) are obtained in shorter intervals. 
We verified this interpretation by fitting Eq.(2) to the same 
subset of RXTE observations used by these authors (more specifically, 
the fit was 
performed on RXTE observations from row 82 to row 114 of Table A1) 
and derived from the best fit ephemeris $n$ =
1.854 $\pm$ 0.003.
However, when these parameters are used to extrapolate the phase shift to the entire 
subset data after the epoch of the glitch claimed by Zhang et al. (2001) 
they produce a systematic deviation of the residuals which increases with the elapsed time, 
as shown in plot of Fig. 5.

\begin{figure}
\centerline{
\vbox{
\psfig{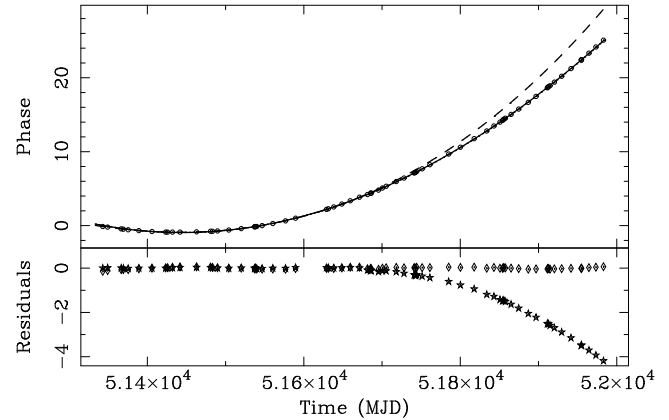}
}}
\caption{Phase residuals obtained extrapolating the ephemerides solution of Zhang et al. 
(2001)
to the entire subset of data after the epoch MJD 51325.
}
\end{figure}

Finally, we note that the agreement of our estimate of braking index
with those derived from ASCA (Hirayama et al. 2002) and BeppoSAX (Mineo et al. 1999)
data can be easily understood on the basis of the rather long time intervals
covered by the observations. They are too sparse to provide a good information on
the timing noise but can give a right evaluation of the mean second derivative.

\begin{acknowledgements}
The authors are very grateful to the referee, R.N.Manchester, for the relevant
comments and suggestions that greatly improved the scientific content of the
paper. This work
has been partially supported by the Italian Space Agency (ASI). 
\end{acknowledgements}

\section*{Appendix}
In this appendix we report a table containing the central epoch of 
the observation 
(column 2), the total duration (column 3),  exposure time (column 4), TOA
and error (column 5 and 6; only for RXTE observations) and the 
frequency with the relative error (column 6,7) for 
each observation analyzed.

\thispagestyle{empty}
\thispagestyle{empty}
\begin{table*}
\begin{center}
\begin{tabular}{llllllll}
\multicolumn{5}{l}{{\bf {Table A1.}} Observation log and frequency measurements }\\
\hline
\hline
   & Tc (MJD) & Tobservation (s) & Texposure (s) & TOA (MJD) & Err ($\mu$s) & Frequency (Hz) &  Err (Hz) \\
\hline
ASCA & & & & & & & \\
\hline
1  & 49151.8084349484  &89463 &30852 & &  & 19.83568828 & 2.4 $\times$ 10$^{-7}$  \\
2  & 49220.5355756515  &94816 &32141 & &  &  19.83456882 & 5.0$\times$ 10$^{-7}$  \\
3  & 49252.0870520244  &64656 &23294 & &  & 19.83405527 & 5.5$\times$ 10$^{-7}$  \\
4  & 49254.4624576429 & 82879 &35572 & &  &  19.83401684 & 1.4$\times$ 10$^{-7}$  \\
5  & 49626.1810950798 & 32480 &11504 & &  &  19.82796873 & 7.0$\times$ 10$^{-7}$   \\
6  & 49628.1308000431 & 24529 &91460 & &  &  19.82793815 & 9.5$\times$ 10$^{-7}$   \\
7  & 49636.3810082158 & 14688 &10550 &  &  & 19.8278009 & 1.9$\times$ 10$^{-6}$  \\
8  & 49637.7172262562 & 14996 &93339 &  &  & 19.8277789 & 1.6$\times$ 10$^{-6}$ \\
9  & 49663.2073326372  &14606 &38299 & &  & 19.8273641 & 3.1$\times$ 10$^{-6}$  \\
10  & 50028.4455501792 &1866565 & 72232 & &  & 19.82142441 & 2.0$\times$ 10$^{-7}$  \\
11  & 50395.3696804385 &2422872 & 90891 & &  & 19.815460211 & 9.2$\times$ 10$^{-8}$ \\
12  & 50759.8892585014 &1638235 & 71682 & &  & 19.80953905 & 1.2$\times$ 10$^{-7}$ \\ 
13  & 51121.2406563684 &2135813 & 85952 & &  & 19.80367343 & 1.5$\times$ 10$^{-7}$  \\ 
14  & 51485.0385356526 &2183033 & 89302 & &  & 19.797770995 &  8.4$\times$ 10$^{-8}$ \\ 
\hline
SAX & & & & & & & \\
\hline
15  & 50382.3304918242 & 71318 & 38347 & &  & 19.81567247 & 2.6$\times$ 10$^{-7}$  \\
16  & 50727.1972055422 & 55830 & 36233 & &  & 19.81007055 & 8.5$\times$ 10$^{-7}$  \\
17  & 51463.4493485718 & 52119 & 23390 & &  & 19.79812490 & 9.6$\times$ 10$^{-7}$  \\
18  & 51465.3785121305 & 39908 & 15803 & &  & 19.79808990 & 3.0$\times$ 10$^{-7}$  \\
19  & 51593.2879147687 & 50285 & 28160 & &  & 19.79601371 & 9.2$\times$ 10$^{-7}$  \\
\hline
RXTE & & & & & & &  \\
\hline
          20 &    50150.2775566760     &  7321  &     4953    &     50150.2775569566      & 915     &    19.8194548     &   7.0$\times$ 10$^{-6}$ \\
          21 &    50221.3528223318     &  6622  &     4921    &     50221.3528227907      & 937     &    19.8182871      &   7.2$\times$ 10$^{-6}$ \\
          22 &    50243.8885962850     &  6698  &     4991    &     50243.8885964943      & 705     &    19.8179247      &   6.1$\times$ 10$^{-6}$ \\
          23 &    50269.1237492340     &  6992  &     4623    &     50269.1237493630      & 1019     &   19.817515     &   1.1$\times$ 10$^{-5}$ \\
          24 &    50296.9179287483     &  7043  &     4991    &     50296.9179291788      & 999     &    19.8170510      &   8.7$\times$ 10$^{-6}$\\
          25 &    50306.7631433992     &  13935 &    11614    &     50306.7631436463      & 233     &    19.8169009      &   1.1$\times$ 10$^{-6}$ \\
          26 &    50332.9466234517     &  6869  &     4926    &     50332.9466237179      & 752     &    19.8164739      &   6.5$\times$ 10$^{-6}$ \\
          27 &    50337.8103624822     &  19359  &    11791   &     50337.8103627079      & 231     &    19.81639654      &   7.6$\times$ 10$^{-7}$ \\
          28 &    50360.5331308203     &  1517  &     1517    &     50360.5331308787      & 1634     &   19.8160378     &   1.6$\times$ 10$^{-6}$ \\
          29 &    50368.4418745432     &  50495  &    31966   &     50368.4418750643      & 193     &    19.81589787      &   2.2$\times$ 10$^{-7}$ \\
          30 &    50390.7548928533     &  12799  &     8719   &     50390.7548930179      & 300     &    19.8155341      &   1.6$\times$ 10$^{-6}$ \\
          31 &    50399.2989191193     &  19839  &     9422   &     50399.2989195186      & 295     &    19.81539563      &   9.3$\times$ 10$^{-7}$\\
          32 &    50404.8100013632     &  13247  &     5870   &     50404.8100015330      & 420     &    19.8153074      &   2.07$\times$ 10$^{-6}$ \\
	  33 &    50423.9342602285     &  110991  &    82523  &     50423.9342607360      & 206     &    19.81499676    &3.7$\times$ 10$^{-7}$\\ 
	  34 &    50425.2694394845     &  115631  &    90731  &     50425.2694396661      & 278     &    19.81497489 &   4.2$\times$ 10$^{-7}$\\                 
          35 &    50437.9618005898     &  71055  & 32169      &     50437.9618006700      & 249     &    19.81476840      &   2.4$\times$ 10$^{-7}$ \\
          36 &    50438.7223552708     &  39727  & 23565      &     50438.7223552489      & 253     &    19.81475603      &   4.5$\times$ 10$^{-7}$ \\
          37 &    50439.2099444506     &  28303 &  10061      &     50439.2099444447      & 407     &    19.81474798      &   8.9$\times$ 10$^{-7}$ \\
          38 &    50439.5834629385     &   17679  & 10703     &     50439.5834632837      & 390     &    19.8147411      &   1.4$\times$ 10$^{-6}$ \\
          39 &    50447.5335356081     &  15871  &     9774   &     50447.5335359889      & 602     &    19.8146102      &   2.6$\times$ 10$^{-6}$ \\
          40 &    50466.4364738898     &   14495  &     9823  &     50466.4364740499      & 567     &    19.8143005      &   2.4$\times$ 10$^{-6}$\\
          41 &    50487.4681557080     &   14911  &     9695  &     50487.4681561890      & 620     &    19.8139646      &   3.2$\times$ 10$^{-6}$ \\
          42 &    50516.4543557125     &   16959  &    10031  &     50516.4543557915      & 662     &    19.8134923      &   2.7$\times$ 10$^{-6}$ \\
          43 &    50528.1082981453     &   17119  &    10511  &     50528.1082981650      & 518     &    19.8133003      &   2.2$\times$ 10$^{-6}$ \\
          44 &    50554.2917732432     &   17119  &    10511  &     50554.2917735571      & 457     &    19.8128801      &   1.9$\times$ 10$^{-6}$ \\
          45 &    50575.1344847706     &   9855  &     7391   &     50575.1344850982      & 576     &    19.8125484      &    3.7$\times$ 10$^{-6}$ \\
          46 &    50617.1578258350     &   8511  &     5135   &     50617.1578260369      & 703     &    19.8118621      &   4.3$\times$ 10$^{-6}$ \\
          47 &    50638.6538020237     &   9215  &     4302   &     50638.6538022807      & 716     &    19.8115068      &   5.2$\times$ 10$^{-6}$ \\
          48 &    50661.8875122125     &   9215  &     4302   &     50661.8875124521      & 563     &    19.8111311      &   2.9$\times$ 10$^{-6}$ \\
          49 &    50680.8301644068     &  13743  &     9087   &     50680.8301645705      & 623     &    19.8108234      &   2.9$\times$ 10$^{-6}$ \\
          50 &    50700.8130593871     &  18991  &     9998   &     50700.8130594987      & 655     &    19.8104987      &    2.2$\times$ 10$^{-6}$ \\
          51 &    50703.1069385382     &  14671  &     9487   &     50703.1069386272      & 670     &    19.8104571      &   2.4$\times$ 10$^{-6}$ \\
          52 &    50710.7365335916     &   17359  &     9710  &     50710.7365336879      & 668     &    19.8103351      &   3.0$\times$ 10$^{-6}$ \\
          53 &    50731.7088303996     &   15983  &    11311  &     50731.7088308387      & 597     &    19.8099983      &   2.5$\times$ 10$^{-6}$ \\
          54 &    50753.7282404459     &   14175  &     9023  &     50753.7282405495      & 628     &    19.8096421      &   2.8$\times$ 10$^{-6}$ \\
          55 &    50775.6616390784     &   14367  &     8335  &     50775.6616392198      & 663     &    19.8092816      &   2.7$\times$ 10$^{-6}$ \\
          56 &    50794.3808143453     &   14303  &     9631  &     50794.3808147188      & 610     &    19.8089757      &   2.7$\times$ 10$^{-6}$ \\
          57 &    50817.4650993146     &   13295  &     9359  &     50817.4651097875      & 679     &    19.8086047      &   3.0$\times$ 10$^{-6}$ \\
          58 &    50838.7500841077     &   13103  &     9871  &     50838.7500840900      & 599     &    19.8082568      &   3.3$\times$ 10$^{-6}$ \\
          59 &    50864.4661456054     &   13103  &     9871  &     50864.4661458375      & 606     &    19.8078396      &   2.5$\times$ 10$^{-6}$ \\
\hline
\end{tabular}
\end{center}
\end{table*}
\newpage
\thispagestyle{empty}
\begin{table*}
\begin{center}
\begin{tabular}{llllllll}
\multicolumn{5}{l}{{\bf {Table A1}} (continued).}\\
\hline
\hline
       & Tc (MJD) & Tobservation (s) & Texposure (s) & TOA (MJD) & Err ($\mu$s) & Frequency (Hz) &  Err (Hz) \\
\hline
          60 &    50884.6064968517     &  18991  &     9758    &     50884.6064972624      & 570     &    19.8075165      &   2.0$\times$ 10$^{-6}$ \\
          61 &    50910.3122764558     &  14784  &     9583    &     50910.3122764554      & 516     &    19.8070939      &   2.4$\times$ 10$^{-6}$\\
          62 &    50939.2524410826     &  15999  &    11263    &     50939.2524415522      & 504     &    19.8066292      &   2.0$\times$ 10$^{-6}$ \\
          63 &    50961.2281108092     &  21087  &     9903    &     50961.2281108402      & 612     &    19.8062677      &   1.6$\times$ 10$^{-6}$ \\
          64 &    50992.1907233621     &  14255  &     9807    &     50992.1907237746      & 540     &    19.8057660      &   2.4$\times$ 10$^{-6}$ \\
          65 &    51014.0147368956     &  17791  &     9967    &     51014.0147372862      & 510     &    19.8054145      &   1.7$\times$ 10$^{-6}$ \\
          66 &    51038.8493459019     &  13343  &     9215    &     51038.8493458773      & 599     &    19.8050143      &   3.0$\times$ 10$^{-6}$ \\
          67 &    51058.7992902631     &  13471  &     9263    &     51058.7992908122      & 768     &    19.8046821      &   3.3$\times$ 10$^{-6}$ \\
          68 &    51085.8265752465     &  9567  &     7167     &     51085.8265753553      & 901     &    19.8042484      &   4.8$\times$ 10$^{-6}$ \\
          69 &    51086.5263870566     &  3839  &     3023     &     51086.5263874268      & 931     &    19.8042625      &   1.8$\times$ 10$^{-5}$ \\
          70 &    51200.6063583254     &  8331  &     5013     &     51200.6063586919      & 569.    &    19.8023902      &   3.7$\times$ 10$^{-6}$ \\
          71 &    51206.7338605356     &  6939  &     5235     &     51206.7338608225      & 455     &    19.8022809      &   4.0$\times$ 10$^{-6}$ \\
          72 &    51220.4579412823     &  7461  &     4927     &     51220.4579415977      & 472     &    19.8020643      &   3.5$\times$ 10$^{-6}$ \\
          73 &    51237.3581160727     &  12119  &    5575     &     51237.3581160859      & 523     &    19.8017889      &   2.9$\times$ 10$^{-6}$\\
          74 &    51256.4577562548     &  7275  &     5101     &     51256.4577562472      & 485     &    19.8014780      &   3.9$\times$ 10$^{-6}$ \\
          75 &    51262.6929378694     &  4917  &     4917     &     51262.6929381281      & 417     &    19.8013909      &   6.1$\times$ 10$^{-6}$ \\
          76 &    51276.3248166093     &  12745  &     4772    &     51276.3248166847      & 500     &    19.8011557      &   2.3$\times$ 10$^{-6}$ \\
          77 &    51294.1430088117     &   8059  &     4919    &     51294.1430087889      & 548    &     19.8008694      &   3.9$\times$ 10$^{-6}$ \\
          78 &    51310.5136331496     &   7286 &  4963        &     51310.5136331665      & 404     &    19.8006079      &   4.1$\times$ 10$^{-6}$ \\
          79 &    51310.8142483174     &   7286 &  4963        &     51310.8142483560      & 491     &    19.8005952      &   3.9$\times$ 10$^{-6}$ \\
          80 &    51311.5969028145     &   4760 & 4760         &     51311.5969030305      & 523     &    19.8005936      &   6.7$\times$ 10$^{-6}$ \\
          81 &    51319.3224291899     &   6599  &     4893    &     51319.3224294960      & 555     &    19.8004542      &   6.4$\times$ 10$^{-6}$ \\
          82 &    51333.0588017849     &   8845  &     5461    &     51333.0588019978      & 594     &    19.8002358      &   3.4$\times$ 10$^{-6}$ \\
          83 &    51342.9771166913     &   8328  &     4895   &     51342.9771167629      & 417    &      19.8000736      &   2.5$\times$ 10$^{-6}$ \\
          84 &    51349.2476702360     &   8259  &     4585   &     51349.2476704655      & 453     &     19.7999788      &   3.0$\times$ 10$^{-6}$ \\
          85 &    51366.0873172747     &   8136   &    5949   &     51366.0873174558      & 443     &     19.7996991      &   3.2$\times$ 10$^{-6}$ \\
          86 &    51366.2957090773     &   7646   &    5943   &     51366.2957094965      & 450     &     19.7996941      &   3.6$\times$ 10$^{-6}$ \\
          87 &    51367.0866244652     &   8135  &    5955    &     51367.0866245036      & 455.     &    19.7996833      &   3.7 $\times$ 10$^{-6}$ \\
          88 &    51369.3537227011     &   6981  &     6003   &     51369.3537228107      & 427.    &     19.7996450      &   3.7$\times$ 10$^{-6}$ \\
          89 &    51375.5511251919     &   6707  &     6707   &     51375.5511254046      & 404     &     19.7995450      &   3.8$\times$ 10$^{-6}$ \\
          90 &    51389.1947032382     &   6861  &     4965   &     51389.1947032309      & 436     &     19.7993291      &   4.0$\times$ 10$^{-6}$ \\
          91 &    51405.9278060073     &   7493  &     4899   &     51405.9278063126      & 443     &     19.7990606      &   3.4$\times$ 10$^{-6}$ \\
          92 &    51422.9600145157     &   7493  &     4899   &     51422.9600149667      & 421     &     19.7987815      &   4.0$\times$ 10$^{-6}$ \\
          93 &    51423.2303145269     &   7175  &   5055     &     51423.2303149586      & 426     &     19.7987702      &   3.6$\times$ 10$^{-6}$ \\
          94 &    51423.8219443136     &   8381  &   6201     &     51423.8219447693      & 384     &     19.7987638     &   2.7$\times$ 10$^{-6}$ \\
          95 &    51425.8291939941     &   10127  &     8005  &     51425.8291944455      & 334     &     19.7987333      &   1.9$\times$ 10$^{-6}$ \\
          96 &    51432.0903727191     &   6151  &     4875   &     51432.0903732235      & 380     &     19.7986228     &   3.7$\times$ 10$^{-6}$ \\
          97 &    51441.7468451228     &   8865  &     5683   &     51441.7468456561      & 380     &     19.7984691     &   2.4$\times$ 10$^{-6}$ \\
          88 &    51462.9139902142     &   18963  &     5107  &     51462.9139906137      & 438     &     19.7981279      &   1.4$\times$ 10$^{-6}$\\
          99 &    51480.3046770383     &   5075   &  5076     &     51480.3046775678      & 434     &     19.7978457      &   6.4$\times$ 10$^{-6}$ \\
          100 &    51481.2968009249     &   5951    &  5027    &     51481.2968014625      & 428     &    19.7978231      &   6.1$\times$ 10$^{-6}$ \\
          101 &    51483.3334558311     &   10179   &   6398   &     51483.3334558612      & 357     &    19.7977987      &   2.2$\times$ 10$^{-6}$ \\
         102 &    51489.6844611122     &   8919  &     6137   &     51489.6844614279      & 423     &     19.7976958      &   2.6$\times$ 10$^{-6}$ \\
         103 &    51504.0707028282     &   18843  &     4896  &     51504.0707030781      & 418     &     19.7974617      &   1.3$\times$ 10$^{-6}$ \\
         104 &    51519.9853613432     &   6697  &     5055   &     51519.9853617755      & 413     &     19.7971995      &   3.6$\times$ 10$^{-6}$ \\
         105 &    51537.0292132529     &   7509  &     5801   &     51537.0292133426      & 391     &     19.7969327      &   3.3$\times$ 10$^{-6}$ \\
         106 &    51537.4694203451     &   8597  &     5263   &     51537.4694203866      & 479     &     19.7969234      &   3.2$\times$ 10$^{-6}$ \\
         107 &    51537.9957500760     &   6211  &     4811   &     51537.9957503765      & 449     &     19.7969143      &   4.9$\times$ 10$^{-6}$ \\
         108 &    51539.7908841811     &  7291  &     4971    &     51539.7908846815      & 395     &     19.7968813      &   3.5$\times$ 10$^{-6}$ \\
         109 &    51546.7092482877     &   10125  &     5153   &     51546.7092487340      & 438     &    19.7967703      &   3.1$\times$ 10$^{-6}$ \\
         110 &    51560.3473796373     &   8443  &     5003    &     51560.3473898280      & 463    &     19.7965521      &   3.3$\times$ 10$^{-6}$ \\
         111 &    51576.6172108713     &    7293  &     5349  &     51576.6172111729      & 451     &     19.7962743      &   4.5$\times$ 10$^{-6}$ \\
         112 &    51589.4437727056     &    8107  &     5453  &     51589.4437728199      & 463     &     19.7960736      &   3.2$\times$ 10$^{-6}$ \\
         113 &    51628.7753839364     &    12515   &   7247  &     51628.7753839364      & 396     &     19.7954385      &   1.8$\times$ 10$^{-6}$ \\
         114 &    51631.2044355514     &    28857  &     8645  &     51631.2044357594      & 329     &    19.79540002      &   6.9$\times$ 10$^{-7}$ \\
         115 &    51638.3028766256     &     19173  &     7803 &     51638.3028766256      & 345     &    19.7952859      &   1.1$\times$ 10$^{-6}$ \\
         116 &    51648.5466267813     &     17863  &     7297  &     51648.5466272009      & 370     &   19.7951216      &   1.4$\times$ 10$^{-6}$ \\
         117 &    51659.3677297694     &     14227  &     7883  &     51659.3677301986      & 349     &   19.7949445      &   1.7$\times$ 10$^{-6}$ \\
         118 &    51671.1320448646     &     12725  &     8077  &     51671.1320448645      & 344     &   19.7947566      &   1.7$\times$ 10$^{-6}$ \\
\hline
\end{tabular}
\end{center}
\end{table*}
\newpage
\thispagestyle{empty}
\begin{table*}
\begin{center}
\begin{tabular}{llllllll}
\multicolumn{5}{l}{{\bf {Table A1}} (continued).}\\
\hline
\hline
       & Tc (MJD) & Tobservation (s) & Texposure (s) & TOA (MJD) & Err ($\mu$s) & Frequency (Hz) &  Err (Hz) \\
\hline
         119 &    51681.3454690879     & 18336  &     8181      &     51681.3454692289      & 420     &   19.7945862      &   1.3$\times$ 10$^{-6}$ \\
         120 &    51684.7731936208     &  4788   &     2699      &     51684.7731939477      & 686     &  19.7945483      &   9.2$\times$ 10$^{-6}$ \\
         121 &    51685.3175718948     &  20279  &     8857    &     51685.3175720315      & 324    &     19.79452255      &   9.6$\times$ 10$^{-7}$ \\
         122 &    51686.2075746935     &  14087  &     7361    &     51686.2075748901      & 364     &    19.7945075      &   1.4$\times$ 10$^{-6}$ \\
         123 &    51695.4979676551     &    8675  &     6585  &     51695.4979680772      & 424     &     19.7943538      &   3.0$\times$ 10$^{-6}$ \\
         124 &    51705.1979969286     &   9037  &     7093   &     51705.1979970477      & 349     &     19.7942000      &   2.4$\times$ 10$^{-6}$ \\
         125 &    51728.0434880623     &    8267  &     5424  &     51728.0434883792      & 659     &     19.7938357      &   4.9$\times$ 10$^{-6}$ \\
         126 &    51740.9588341276     &    14019  &     8041  &     51740.9588343721      & 586     &    19.7936227      &   2.5$\times$ 10$^{-6}$ \\
         127 &    51741.7215067928     &     9785  &     7727  &     51741.7215072644      & 355     &    19.7936115      &   2.1$\times$ 10$^{-6}$ \\
         128 &    51742.8596909607     &     9907  &     7733 &     51742.8596914658      & 333     &     19.7935918      &   2.0$\times$ 10$^{-6}$ \\
         129 &    51744.2896447745     &   13735  &     7827   &     51744.2896451619      & 333     &    19.7935685      &   1.6$\times$ 10$^{-6}$ \\
         130 &    51751.1810596555     &    9819  &     8123  &     51751.1810597295      & 331     &     19.7934563      &   2.1$\times$ 10$^{-6}$ \\
         131 &    51761.1185637994     &   13057  &     7771   &     51761.1185642818      & 322     &    19.7932910      &   1.5$\times$ 10$^{-6}$ \\
         132 &    51785.0980336962     &   24995  &     9013   &     51785.0980341569      & 307     &    19.79290630      &   7.2$\times$ 10$^{-7}$ \\
         133 &    51800.1416986946     &   8874 &        7731   &     51800.1416989359      & 341     &   19.7926609      &   2.3$\times$ 10$^{-6}$ \\
         134 &    51817.7362062067     &   7231&        5409   &     51817.7362064879      & 683     &    19.7923712      &   7.2$\times$ 10$^{-6}$ \\
         135 &    51833.7908365378     &   8432 &        5217   &     51833.7908366692      & 822     &   19.7921206      &   5.0$\times$ 10$^{-6}$ \\
         136 &    51842.9466249313     &   18462&        8713   &     51842.9466252357      & 509     &   19.7919684     &   1.9$\times$ 10$^{-6}$ \\
         137 &    51850.7533332734     &   12975&        8503   &     51850.7533333821      & 520     &   19.7918429      &   2.3$\times$ 10$^{-6}$ \\
         138 &    51854.1343244782     &    12300&        8261  &     51854.1343246987      & 337     &   19.7917833      &   1.8$\times$ 10$^{-6}$ \\
         139 &    51854.7806751278     &   9334&        7205   &     51854.7806753799      & 389     &    19.7917757      &   2.1$\times$ 10$^{-6}$ \\
         140 &    51855.7761402855     &    9391&        7379  &     51855.7761404915      & 359     &    19.7917586      &   2.2$\times$ 10$^{-6}$ \\
         141 &    51857.6643855983     &   14666 &        7715   &     51857.6643857372      & 372     &  19.7917283      &   1.7$\times$ 10$^{-6}$ \\
         142 &    51864.8967614676     &   3471&        7215   &     51864.8967616390      & 467     &    19.791594      &   1.0$\times$ 10$^{-5}$ \\
         143 &    51886.9135988688     &   10217&        8413   &     51886.9135993413      & 460     &   19.7912528      &   2.5$\times$ 10$^{-6}$ \\
         144 &    51896.8522715505     &   8927&        5353   &     51896.8522718532      & 382     &    19.7910941      &   2.4$\times$ 10$^{-6}$ \\
         145 &    51911.2433041537     &    19227&        6145  &     51911.2433043693      & 415     &   19.7908604     &   1.2$\times$ 10$^{-6}$ \\
         146 &    51911.8487562932     &   9263&        7181   &     51911.8487567141      & 397     &    19.7908517      &   2.5$\times$ 10$^{-6}$ \\
         147 &    51912.8241077159     &   12570&        7301   &     51912.8241078523      & 388     &   19.7908334      &   2.0$\times$ 10$^{-6}$ \\
         148 &    51914.6370942449     &   8584&        6203   &     51914.6370947508      & 355     &    19.7908040     &   2.3$\times$ 10$^{-6}$ \\
         149 &    51920.2291815557     &   13749&        5932   &     51920.2291819350      & 444     &   19.7907147      &   1.9$\times$ 10$^{-6}$ \\
         150 &    51929.5870831782     &   14465&        7573   &     51929.5870831607      & 378     &   19.7905631      &   1.6$\times$ 10$^{-6}$ \\
         151 &    51941.4093028956     &   13340&        6081   &     51941.4093032091      & 464     &   19.7903715      &   2.1$\times$ 10$^{-6}$ \\
         152 &    51954.5931647284     &   12499&        7201   &     51954.5931647444      & 810     &   19.790178      &   4.3$\times$ 10$^{-5}$ \\
         153 &    51954.7239382431     &   12499&        7201   &     51954.7239384635      & 538     &   19.7901562      &   2.0$\times$ 10$^{-6}$ \\
         154 &    51954.9246680830     &   12373&        7391   &     51954.9246684580      & 596     &   19.7901536      &   2.6$\times$ 10$^{-6}$ \\
         155 &    51964.4474052025     &   8564&        8118   &     51964.4474055636      & 387     &    19.7900000      &   2.5$\times$ 10$^{-6}$ \\
         156 &    51973.2333393971     &    24758&        9000  &     51973.2333393914      & 501     &   19.789861      &   1.0$\times$ 10$^{-5}$ \\
         157 &    51982.9725223272     &    25961&        7904  &     51982.9725226691      & 371     &   19.78969766      &   7.8$\times$ 10$^{-7}$ \\
\hline
\end{tabular}
\end{center}
\end{table*}

\end{document}